\documentclass[pre,twocolumn,superscriptaddress,floatfix]{revtex4}
\usepackage{listings}
\usepackage{fancyvrb}
\usepackage{graphicx}
\usepackage{latexsym}
\usepackage{bm}
\usepackage{natbib}
\usepackage{dcolumn}
\usepackage{amsmath}
\usepackage{amstext}
\usepackage[english]{babel}
\usepackage{flafter}
\usepackage{url}
\usepackage{multirow}
\usepackage{float}
\usepackage[usenames]{color}
\newfloat{figtable}{t}{lop}
\floatname{figtable}{Table}

\begin{document}

\title{Editorial}
\author{Santo Fortunato}
\affiliation{Department of Biomedical Engineering and Computational
  Science, School of Science, Aalto University, P.O. Box 12200, FI-00076, Espoo, Finland}
\author{Michael Macy}
\affiliation{Department of
Sociology, Cornell University, Ithaca, New York, USA}
\author{Sidney Redner}
\affiliation{Center for Polymer Studies and Department of Physics, Boston University, Boston, Massachusetts, 02215 USA}

\maketitle

Statistical physics has proven fruitful for the investigation of the
collective dynamics of complex systems, including systems that lie beyond the
scope of traditional physics. By leveraging behavioral regularities at the
global level, such as averages and distributions, statistical physics can be
used to analyze systems with large numbers of components whose individual
behavior is highly idiosyncratic This regularity also occurs when the
fundamental constituents are more complex than atoms or molecules.  In the
XVIII$^{\rm th}$ century it was first noticed that events such as the number
of births, deaths, suicides, etc., tended to be stable in a given
geographical area, as long as the observation period is not too long.  This
stability was surprising because these events are generally unpredictable
individually. These empirical regularities motivated Maxwell and Boltzmann to
propose a statistical approach to understand the physics of many-particle
systems, ultimately leading to the foundations of statistical
mechanics~\cite{ball04}.  Related observations also convinced scholars of the
period that precise quantitative laws, like those of physics, also existed
for social phenomena, in spite of the apparently erratic behavior of
individuals.  For example, Immanuel Kant, in his 1784 essay {\it On History}
refers to universal laws that ``{\it however obscure their causes, [permit]
  us to hope that if we attend to the play of freedom of human will in the
  large, we may be able to discern a regular movement in it, and that what
  seems complex and chaotic in the single individual may be seen from the
  standpoint of the human race as a whole to be a steady and progressive
  though slow evolution of its original endowment}''.

This conviction that there should exist a quantitative theory of social
phenomena was shared by many noted scholars, such as the Marquis de
Condorcet, Auguste Compte (credited as the father of sociology), Adolphe
Quetelet, John Stuart Mill, Henry Thomas Buckle, to name a few.  Yet more
than two centuries after the empirical discovery of the stability of social
aggregates, we still lack a social science equivalent of Newton's laws for
the motion. There are two main reasons. The first is the relative dearth of
empirical observations compared to those in physics.  Robert C. Merton, a
Nobel Laureate in Economics, pointed out that Newton's laws were the result
of a many-decade process of careful observations by Danish astronomer Tycho
Brahe, which were subsequently summarized in mathematical form by Johannes
Kepler.  The regularity of Kepler's observations inspired Newton's work.
Simply put, the absence of a theory of society is not that we are still
awaiting the social counterparts of Newton and Einstein, Merton concludes,
but rather we have not yet found our Brahe and Kepler.

Additionally, humans are complex, while elementary particles are simple.
Since the early experiments by Galileo, physics has focused on elementary
objects and their interactions whose simplicity made it possible to derive
the basic laws of particle motion, then electromagnetism, and finally the
laws of nuclear and subnuclear interactions. By means of powerful
mathematical tools, like probability theory, it became possible to recast
these laws of elementary interactions in few-particle systems into
statistical laws that account for the behavior of many-particle systems.


When we attempt to apply this same minimalist modeling to social phenomena,
we face several obstacles:
\begin{itemize}
\item[]\textit{Heterogeneity.}  While the constituents of physical systems
  are {\it homogeneous}, those of social systems are not.  As the Physics
  Nobel Laureate Wolfgang Pauli said, ``{\it in physics we can assume that
    every electron is identical, while social scientists do not have this
    luxury}''.  While the interactions between electrons under identical conditions are the same, this reproducibility is not true for people, not
  only because people are unique, but also because humans may {\it learn} and
  {\it adapt}~\cite{buchanan07}.  Thus the same two individuals may react
  differently in repeated identical situations because of accumulated
  knowledge.  In fact, since history is played only once, we generally have
  distorted perceptions of social behavior, overemphasizing what actually
  happened, to the detriment of what could have happened~\cite{watts11}.

\item[]\textit{Emergence.} The behavior of a population does not
  necessarily imply aggregation of individual behaviors.  For example, highly
  segregated residential neighborhoods can emerge in populations of
  individuals that not only tolerate diversity but even prefer it.
  Similarly, the properties of a collective can vary over time even if those
  same properties are constant over time at the individual level.

\item[]\textit{Reification.} Collective decision making does not occur
  through cognitive processes analogous to decision making by
  individuals---i.e., there is no ``group mind'' and properties of individual
  actors, like intentionality, purpose, regret, or anger, are not properties
  of collective actors.

\item[]\textit{Correlation vs.\ Causality.} Statistical regularities are
  useful to identify possible individual-level explanations for population
  dynamics.  However, such correlations do not imply causation.  In the
  absence of general theories, confidence in having identified the
  underlying causal mechanism usually depends on confirmation based on controlled
  experiments.

\end{itemize}

These differences between humans and atoms illustrate the difficulties that
confront efforts to create a physics-like science of social life. 
Even if social scientists had been able to match the successes of physicists in
collecting observations of social phenomena, they would have probably failed to
extract elementary laws.

However, lessons from critical phenomena give hope that statistical physics
can help quantify the dynamics of large groups of individuals.  In many phase
transitions, the large-scale behavior of a many-particle system is
independent of particulate details and their microscopic interactions; only a
few basic features are relevant.  Thus systems that cannot be fully
characterized at the individual level might still display recognizable
patterns in the aggregate, if the number of constituents is sufficiently
large. This applies not only to the collective behavior of atoms but to human
behavior as well.

Building on this analogy, a statistical physics of society, or
``sociophysics''~\cite{castellano09} seems a feasible goal.  In recent years,
many simple models, inspired by statistical physics, have been introduced to
account for social phenomena such as opinion formation, cultural and language
evolution, collective action, crowd behavior, Web user behavior, marketing,
financial panics, political polarization, and neighborhood segregation.
Although these idealized models have attracted growing interest, their
relevance for understanding real-world social dynamics is highly
controversial.  Models derived from physics are often highly stylized,
and are typically based
on simplifying assumptions about the mechanisms driving both individual and
collective behavior.  These simplifications can be useful in thought
experiments designed to identify the macro-level implications of a set of
behavioral and environmental assumptions but they can also compromise the
ability to explain real-world outcomes. And until recently, efforts to
introduce more empirically plausible assumptions had to confront the absence of solid
empirical data. 

That limitation on data is changing rapidly. Social systems with many
individuals have now become experimentally observable, due to the rapid
increase in computational power to process big data at very low cost, and the
growing use of the Web. This is both an increasingly indispensable medium of
interaction, as well as a vast repository of the digital traces of those
interactions and a platform to carry out controlled experiments involving
large numbers of participants.  Due to the increasing availability of data,
other disciplines, like computer science, have entered this line of endeavor
under the rubric of {\it computational social
  science}~\cite{lazer09,giles12}.  Sociophysics models cannot avoid
validation any more.


This special issue on ``The Application of Statistical Mechanics to Social
Phenomena'' represents our attempt to present and confront several major
challenges in sociophysics.  Our first goal is very modest --- to promote
inter-disciplinary cross-fertilization, by making statistical physicists more
aware of the perspectives and long-standing contributions of quantitative
social scientists and by providing quantitative social scientists a glimpse
into the modeling approaches of statistical physics.  Both fields can
certainly learn from the other and we hope that this issue will help in this
endeavor, both in providing exposure to new ideas across fields and also
making people aware of some of the active researchers in each field.

A second goal is to expose a representative range of social phenomena that
are currently under study by social scientists and by physicists, and to
illustrate the perspectives and techniques that are used by current practitioners
in their respective subfields.  We anticipate that some researchers are only
vaguely aware of very relevant investigations that appear in the literature
of complementary fields.  We therefore hope that this volume helps foster, in
some small measure, better communication between the
physical and social sciences.

Let us now provide an overview of the contributions to this special issue:

\noindent{\bf Volume I}

In his very entertaining and idiosyncratic style, D. Stauffer begins this
volume with a brief review of several classic models of social dynamics that
have captured the attention of physicists, including the Schelling model of
segregation, idealized models of opinion dynamics, and the dynamics of
scientific citations.  He also raises the fundamental question---does
sociophysics make sense?---that we hope will be partially answered in the
positive by this volume.

The following section discusses a variety of stylized models of opinion
evolution.  For physicists, an important starting point is the voter model,
in which each individual is endowed with one of two discrete opinion states.
The update step is simplicity itself: a random individual is picked and
updates his opinion by adopting the state of one of his neighbors.  This
update step is repeated until consensus necessarily occurs in a finite
system.  This description of people as identical atomistic elements is
appealing to statistical physicists and has spawned much research.  The
articles on opinion dynamics in this volume involve atomistic descriptions
that incorporate either different modeling perspectives or certain aspects of social
reality.

For example the article by Lanchier presents some rigorous results on the
evolution of the so-called (Galam) majority-rule model, in which the dynamics
operates on all the members of randomly chosen groups.  All members of the
group adopt the local majority state and repeated updates of this type
ultimately lead to consensus.  The article by Galam himself
discusses new collective features when there are three competing voting
states.  Some of these features are amplified in the article by Mobilia on
the competition between a persuasive majority and a committed minority in a
three-state voter model.  Another important aspect of opinion dynamics is the
very real possibility of disagreement.  Some features of this attribute are
discussed in the article by Li et al.\ on the so-called non-consensus model in
which the interactions permit the coexistence of two disagreeing states, rather
than a monotonic evolution toward consensus.  The role of the interaction
range in the voter model is explored in the article by Pastor-Satorras et al.\
Here a voter changes opinion to a given opinion state only if q of its
neighbors are in that state.  An important aspect of this class of models is
that it breaks magnetization conservation (the density difference between
voters of opposite states).  The distinction between conserved and
non-conserved voter models has been an active area of investigation.

Instead of one voter simply adopting the state of another, two individuals
may choose to compromise and average their opinions.  This appealing model of
compromise, first introduced by Deffuant et al.\ and by Hegselmann and Krause
has spawned much research into understanding its rich dynamics.  The article
by Carro et al.\ investigates the role of noise and initial conditions on the
opinion states of all individuals in the long-time limit.  The article
by van
Brecht et al.\ explores the compromise model under the combined effects of
attractive and repulsive interactions, and with pairwise interactions defined
on the Erd\"os-R\'enyi random graph.  As the density of pairwise interactions is
varied the stability of the long-time opinion equilibrium changes
drastically.  The article by Nyczka and Sznajd-Weron explores the influence
of microscopic features of social reality on opinion evolution models, such as
the topology of local interaction, the influence of external interactions,
the relative role of conformity/anti-conformity, etc.  Under the rubric of
opinion dynamics, the contribution by Torney et al.\ introduces a voter-like
model that is intended as an idealized model of the movement of individual
animals that are part of a large group.  This model incorporates the notions
of a ``correct'' choice for one of the two opinion states and ``leaders,''
namely, individuals who ``know'' the correct choice and cannot be persuaded to
move from this opinion state. Loreto et al.\ explore the interplay between
disagreement and external information in opinion dynamics. They find that
moderate messages have better success compared to more extreme information.
Models of collective action are similar to models of consensus
formation in showing how 
individual behavior can often have surprising consequences for the
population dynamics. 
Centola uses formal models to challenge the widely held
assumption that free riding reduces the social efficiency of
collective action.  On the contrary, self-reinforcing incentives to
participate in collective action reduce free riding but make outcomes
vulnerable to perturbation.  Izquierdo and Izquierdo discuss the
applicability of the mean-field approximation to stochastic processes
expressable as Markov chains, like many computer models. They derive
results from stochastic approximation theory on the relation between
computer simulations results for simple stochastic models and their
expected deterministic behaviors in the mean-field approximation.

Language dynamics is a subject for which statistical physics ideas seem to be
gaining some traction.  There are many intriguing regularities in language
data, such as word frequencies, the number of distinct words as a function of
the corpus size, etc., that are well-suited for statistical analyses.  In
this vein, the article by Altmann and Motter uses word-frequency dynamics to
identify social trends in the dynamics of individuals themselves.  The
article by Fujie et al.\ extends the Abrams-Strogatz model for the competition
between two languages in a population in which there are more than two
languages.  In the original Abrams-Strogatz model, the lower-status language
always becomes extinct, but language coexistence can arise if more than two
languages are in competition.

With the increased availability of pervasive location data that is gleaned from
cell phone calls, credit-card transactions, and smart phone applications, the
possibility of modeling urban human mobility patterns has moved from being
unattainable to reality.  The contribution by Hasan et al.\ investigates the
individual mobility patterns from smart subway fare transactions in London
and the insights that can be gained from these patterns.

Everyone is interested in popularity.  The provocative article by Simkin and
Roychowdhury explores the relation between popularity and achievement in data
on WWI aces.  These authors conclude that there is a power-law distribution
for individual fame.  This result seems to apply to many other human
activities.  Another aspect of popularity is the notion of "attention".  We
are so heavily inundated with information and advertising, that what now
matters is how much attention we actually pay to the myriad of sensory inputs
that we experience.  The article by Huberman explores the relative efficacies
of popularity and novelty in determining how to prioritize information.  An
important measure of popularity in the sciences is scientific citations.  We
all care about how often our work is cited.  The contribution by Golosovsky
and Solomon analyzes the citation network of all of {\it Physical Review} and
argues that the citation dynamics is governed by superlinear preferential
attachment.  They also find evidence of very different citation dynamics of
poorly cited papers (which quickly disappear) and highly-cited paper (which
seem to be immortal on the time scale of {\it Physical Review}).

An important aspect of social dynamics is the spreading of a disease or a
rumor through a population.  The most well-studied epidemic models are the
SIS (susceptible-infected-susceptible) and SIR
(susceptible-infected-recovered) models.  In the former, a susceptible (S)
individual is exposed and becomes infected by interacting with an I, but an I
eventually recovers and becomes an S again; such a situation describes the
common cold, where one can be infected many times.  In the SIR model, a
recovered individual becomes immune and is given the label R for recovered.
The paper by Tunc et al.\ explores the ramifications of individuals breaking
their social contacts in response to an epidemic.  In particular, the
location of the epidemic threshold was found to depend on the rate at which
an individual deactivates links with acquaintances who have fallen ill.  The
article by Lund et al.\ investigates the effect of heterogeneous city
populations on epidemic spreading within the SIS model.  Here, the epidemic
dynamics within a city is treated in the mean-field approximation, and travel
between cities then couples these dynamical systems.

Classic and well studied
rumor-spreading models have been developed that incorporate three types of
individuals: spreaders, stiflers (those that are aware of the rumor but no
longer wish to spread it), and those unaware of the rumor.  The evolution of
these three classes of individuals is reminiscent of the SIR epidemic model.
The paper by Borge-Holthoefer et al.\ exploits currently available online data
to expose the inadequacies of this classic model and points the way for more
realistic modeling of spreading phenomena.

\noindent{\bf Volume II}

While diffusion and social influence are often assumed to involve
processes like persuasion and imitation, influence can also be
negative, in which the recipient responds by differentiating from the
influencer.  The contribution by Johnson et al. explores a simple way in
which individual cultural and behavioral traits (e.g., ethnicity) can be
incorporated in a model without compromising tractability.  This
heterogeneous model shows the same power-law distribution of event severity
as in models with homogeneous agents, suggesting that this power-law behavior
is universal.  Another forum for the appearance and resolution of conflict is
in Wikipedia.  While it is hard to imagine how such an anarchistic and
collaborative encyclopedia can work, it has been extraordinarily successful
and useful.  The contribution of Yasseri and Kert\'esz investigates the
dynamics of this collaborative environment and models how conflicts arise as
the document develops and how they get resolved.

In the arena of human endeavor, we are (it is hoped) all trying to better
ourselves.  The contribution of Bardoscia et al.\ represents an appealing
attempt to formalize this improvement process through a "social climbing
game".  The basic idea is that individuals change their links in a social
network to preferentially link to important (high-degree) individuals and to
also maximize their number of links.  This model undergoes a phase transition
to a highly hierarchical state as a social temperature---a measure of the
gain in utility in an advantageous update---decreases below a critical value.
The recent availability of on-line sports statistics provide another rich
data source with which to apply ideas from statistical physics and quantify
social hierarchies.  The article by Ben-Naim et al.\ presents a simple model
of competitions in which the only parameter is the probability that the
weaker team (as measured by current win/loss records) upsets the stronger
team.  The model is applied to various sports competitions, where it is shown
that single-elimination tournaments are unfair because there is an
appreciable probability that a strong team is quickly eliminated or a weak
team succeeds.  A tournament schedule is devised that is both fair and
efficient, in that relatively few games are needed to determine the strongest
team.

Economic systems historically have represented one of the main arenas for the
application of physics modeling.  In the paper by Weisbusch, the author
tackles a very important question: will economic growth lead in the future to
more or less inequality among world regions?  The author uses a spatial model
whose basic assumptions are the limitation on resource influx and the
tendency to concentrate production on richest producers.  The conclusion is
that the inequality will persist.  Gordon et al.\ use the Random Field Ising
Model and the Schelling model to describe the influence of social
interactions in market situations, which often lead to a choice of
buying/selling that would not be taken otherwise (bandwagon effect).  They
find that the price originally posted by the seller is already close to the
critical value beyond which there will be little response by other customers,
so prices of goods with bandwagon effects do not increase despite their
success, as observed in real market scenarios.  Koskinen and Lomi find that
the standard "gravity" model of international trade describes the network of
foreign direct investment relations in the international electricity industry
quite well.  Toscani et al.\ adopt Boltzmann-like equations for the evolution
in time of the density of wealth in a market composed by many agents. Some of
the solutions of these equations develop Pareto tails.  Bouchaud shows that
the Random Field Ising Model provides a unifying framework to describe the
appearance of sudden crises and ruptures in many collective socio-economic
systems.  Cascades of events leading to such extreme events are generated
either by herding, i.e., reference to peers, or trending, i.e., reference to
the past.  Time scales are crucial, as they may sometimes be so long that
social systems cannot reach optimal equilibria.  Rapisarda et al.\ highlight
the importance of random strategies in the social sciences, focusing on the
choice of political representatives and financial trading.

Evolutionary game theory, originally developed within biological contexts,
has, in recent times, attracted social scientists, economists, and
philosophers. In addition to biological evolution it is natural to consider
social evolution. In our collection there are four contributions in this
area. The paper by Hernandez and Zanette proposes an evolutionary framework
for the "Colonel Blotto" Game, a game of how to distribute limited resources
among different channels (or battlefields) so as to maximize the probability
of long-term success. The authors show that optimal strategies emerge as
equilibria of the dynamics, and that this result is robust in generalizations
of the model.  Fu and Nowak study the influence of migration in the evolution
of cooperation within structured populations.  In a finite island model,
where individuals play games with others on the same site, they find that
global migration can lead to stronger spatial effects than local migration
for a wide range of the model parameters.  Christoforou et al.\ study the
evolution of cooperation by considering the context of crowd computing, which
has become very popular in the last few years. They show that the system
converges to the exact solution of a given assignment in many cases, even
when many participants are defectors.  In the paper by Short et al., ties
among criminals are introduced in an adversarial evolutionary game in which
crimes are perpetrated and criminals may be convicted if enough witnesses
testify against them.  The presence of these ties (or "sacred values")
enhances the persistence of crime even in fairly peaceful communities.

The structure of complex networks and dynamics on these networks have become
part of the backbone of complexity science. Social networks in particular
have been and are currently being intensely investigated, in large part for
their practical importance as communication and information media (e.g.,
Facebook).  T\'oth et al.\ present a study on the detection of communities in
networks, a recent and very hot topic, consisting in the identification of
cohesive groups of nodes sharing but a few links with the rest of the
network. The authors focus on the "clique percolation" method, which is based
on exploring the network by following sets of overlapping cliques. The size
of the cliques is traditionally chosen such that the system lies just below
the critical point for the percolation of the cliques. The paper justifies
this assumption by showing that the quality of the partition is higher in
this case, according to a quality measure of overlapping community
structures.  Klemm asks how searchable networks are, i.e., how easily one can
reach high centrality nodes by local search. By introducing a measure called
smoothness, he shows that many networks have close to optimal searchability
for eigenvector centrality, while other measures, like degree and betweenness,
are harder to track. Musumeci et al.\ present a method to reconstruct complex
networks starting from a partial information on their structure. Starting
from information about some intrinsic fitness of the known nodes, the missing
links are detected by calibrating a fitness model on the known links. Tests
on synthetic networks and on the World Trade Web show that a faithful
reconstruction is possible starting from as little as 10\% of the nodes. Kaski
et al.\ focus on egocentric networks, i.e., the networks of neighbors of a
source node (ego). They investigate the difference between the best
connection of the ego and the second-best through an analysis of mobile phone
communication networks. They find that the ratio of the number of calls of
the ego to his most frequent and second most frequent contacts, above a
threshold, is a good index for the reliability of the identification of an
individual's closest acquaintance.
The contribution by Altshuler et al. develops a methodology to identify
anomalous informational signals in social networks by monitoring cell phone
calls that pass through the edges around hubs.  This technique was found to
efficiently identify emergencies in cell phone call data.

Latora et al. explore the relationship between social capital and
structural network measures.  In particular, the clustering coefficient and
the effective size of the neighborhood of an actor in a social network,
indicating the number of contacts that can only be reached by the actor
through a direct link, have been both proposed as suitable measures for
social capital. The authors show that these two variables are actually
dependent on each other and introduce a new variable that mediates between
the two, quantifying the role of bridges that mediate between otherwise
disconnected nodes. Schweitzer et al.\ present an
analysis of systemic risk on network systems, determined by cascading
failures of macroscopic size. They focus on a system of agents that fail if
their load exceeds a given threshold. The authors find that if the threshold
distribution is a power law, finite cascades are still possible, while random
fluctuations can trigger full cascades. In general, the size of the shock
initiating the cascade is not as important as the network topology and the
threshold distribution.

\end{document}